**An Experimental Study of Segregation Mechanisms**

Milena Tsvetkova[a,1] , Olof Nilsson[b], Camilla Öhman[b], Lovisa Sumpter[c], and David Sumpter[b]

[a] Oxford Internet Institute, University of Oxford, Oxford OX1 3JS, United Kingdom

[b] Department of Mathematics, Uppsala University, 75105 Uppsala, Sweden

[c] School of Education, Health and Social Studies, Dalarna University, 79188 Falun, Sweden

[1] To whom correspondence should be addressed. Address: Oxford Internet Institute, 1 St Giles, Oxford OX1 3JS, United Kingdom. Telephone: +44 1865 612330. E-mail: milena.tsvetkova@oii.ox.ac.uk

**Abstract**

Segregation is widespread in all realms of human society. Several influential studies have argued that intolerance is not a prerequisite for a segregated society, and that segregation can arise even when people generally prefer diversity. We investigated this paradox experimentally, by letting groups of high-school students play four different real-time interactive games. Incentives for neighbor similarity produced segregation, but incentives for neighbor dissimilarity and neighborhood diversity prevented it. The participants continued to move while their game scores were below optimal, but their individual moves did not consistently take them to the best alternative position. These small differences between human and simulated agents produced different segregation patterns than previously predicted, thus challenging conclusions about segregation arising from these models.

**Introduction**

Even if people are generally tolerant, they can end up living in a segregated society if they have weak preferences to be among similar others. This was the counter-intuitive conclusion derived from Schelling's model of segregation (1, 2), one of the most influential models of tipping dynamics and unintended consequences (3-5). The most common version of the model is a simulation in which agents of two different colors are placed on a grid. On each time step of the simulation, the agents move position depending on the color of their eight nearest neighbors. For example, the agents move if less than 50% of their neighbors have the same color as themselves (Figure 1A). Even when everyone has these relatively mild preferences for similar neighbors and no one prefers a segregated local neighborhood, the model dynamics result in extreme segregation at the population level (Figure 1E).

Subsequent theoretical research has shown that even when individuals actively seek diversity, segregation may still be a likely outcome (6-8). These models assume that the larger the proportion of same-color neighbors, the less satisfied the agents are and the more likely they are to move (as in, for example, Figure 1C). Despite diversity providing higher utility, the simulations predict a surprisingly high level of segregation (Figure 1G).

When considering questions of public policy, both economists and sociologists have emphasized the importance of these results. For example, (8) conclude that "segregation can emerge…even among multiculturalists who actively seek diversity, so long as they are also sensitive to small changes in ethnic composition." Moreover, (7) suggest that the "welfare effect of educating people to have preferences for integration might be adverse" because the "segregated outcome will be unsatisfying for the majority of people." The practical implications are clear: if the theoretical predictions are true, public policies that promote openness and tolerance are futile because they cannot improve integration.

Despite their importance for policy, the predictions of the segregation models have not been extensively tested empirically before. Most of the previous work has investigated what kind of preferences for neighborhood composition individuals from different racial and ethnic groups hold and what level of segregation these preferences produce if plugged into the models (4, 10-12). The question whether particular preferences indeed cause the predicted outcomes has received less attention. Answering this question with observational data is extremely difficult since there are multiple correlated factors that influence a person's moving decision (13). For example, in a residential setting, the decision to move a house is affected not only by the ethnic composition of the current and future neighborhood, but also by the household's income, the price of real estate, proximity to the workplace, etc. (4). In a conference room or a lecture hall, the decision to pick a particular seat may be influenced by the characteristics of the seating neighbors but may also have to do with when one enters the room or one's personality type (14).

The best method to isolate the effect of a behavioral factor is to conduct a controlled experiment (15-17). So far there has been only one small-scale experimental test of a one-dimensional version of the Schelling model (i.e. with a utility function as in Figure 1A) (18). We therefore designed, conducted, and analyzed an experiment to test the predictions of the two-dimensional segregation model for four different utility functions, representing different preferences for local similarity, diversity, and difference. For simplicity, we refer to the corresponding games as "Same," "Same and Diverse," "Diverse," and "Same or Different" (Figure 1A-D).

**Materials and Methods**

We conducted the experiments with 20 high school classes in Sweden. The classes had between 13 and 25 students, providing a total of 399 participants, aged between 16 and 19. To recruit the classes, we identified mathematics teachers in three different regions of Sweden (east coast, middle, and west coast) and contacted them with the offer to lead an hour-long interactive demonstration on mathematical modeling of social processes during one of their class sessions. The experimental sessions were conducted at the beginning of the demonstrations and introduced as a game with no reference made to segregation. Each participant was given a surf tablet to use as a game controller and assigned a number and a color for their avatar (Figure S2). The game was projected on a shared screen at the front of the classroom (Figure S3). The game field was a six-by-six square grid on which avatars could be moved up, down, left, or right. Participants played four games with scoring rules equivalent to the four utility functions (Figure 1A-D), in randomized order. Participants could observe their current score in real time throughout the game, but only the score at the end of the game counted. Each game was stopped when all participants indicated that they did not want to move any further, or after 2 minutes of play.

An important novelty of our experimental design is that it was performed in a naturally occurring social environment. Most game experiments reported in the literature place subjects in isolated booths in computer laboratories, do not allow communication, and incentivize performance with monetary payments. By running our games in classrooms, using tablets, and having interactions occur on a shared screen, we gave participants the opportunity to interact naturally, while still being able to record all of their game actions. We allowed participants to communicate with each other but this did not significantly de-anonymize their interactions (Figure S4). We did not offer monetary incentives but the games were very dynamic (3.8 moves per second on average) and caused a lot of excitement and exclamations during game play. We provided a social incentive to perform well by informing participants in advance that their total score from the four games would be projected on the screen at the end of the experiment.

**Results**

In the Same game, participants produced slightly higher levels of segregation and average scores than the model predictions (Figure 2A, 3A). In 17 out of the 20 trials the participants separated in two distinct groups at either side or corner of the grid (as in Figure 1I). The Same game was the easiest game for the participants to complete: nearly all groups converged to the end-game segregation and average scores in less than 30 seconds of play (Figure S5). This was because it quickly became apparent which side "belonged" to which color.

For the Diverse game, the segregation was low and similar to the model predictions. The participants continued to move for the entire two minutes of the experiment (Figure S5) but they eventually achieved the expected good levels of integration (Figure 2B) and high average scores (Figure 3B).

The experimental results for the Same and Diverse game did not reveal marked levels of segregation, contradicting model predictions. In most cases, participants actually achieved a good degree of integration (Figure 2C). Yet, their scores were not always lower than the simulated agents' scores (Figure 3C). The high level of segregation in the simulation was due to the fact that a small number of agents quickly formed "a frontline" that offered the ideal mix of own- and other-color neighbors; the rest of the agents were doomed to less-than-optimal positions and logically, they chose the own-color side of the borderline (Figure 1G). In contrast, experiment participants achieved a more uniform mix (Figure 1K), similar to the pattern in the Diverse game (Figure 1J).

The experimental results for the Same or Different game also deviated from the model. The students tended to have higher levels of segregation (Figure 2D). Further, based on the average scores, it is clear that the participants did worse in the game than the simulation agents (Figure 3D). A priori, we expected that the participants would recognize that the "same" strategy was much easier to coordinate on and hence, they would converge to high levels of segregation, as they easily did in the Same game. During the experiments, students made repeated references to such a strategy, shouting things like "all the yellows up to the top." Four groups attempted to carry out this approach but it took only a few contrarians to upset the pattern (Figure S5D). However, no group managed to achieve the simpler one-color-neighborhood solution. The students also made suggestions about checkerboard solutions like "we should alternate….blue then yellow" or "we'll stand in groups of four two of each with one square distance between." Three groups managed to coordinate (or nearly coordinate) to create these more difficult checkerboard solutions. The rest of the groups failed to get close to a mutually beneficial configuration within the allotted time.

The vast majority of moves made by the participants were consistent with an attempt to maximize the utility functions provided to them. Figure 4 shows the average latency until a move is made as a function of neighbor types for the four games. Movement patterns differed greatly between games, but were consistent across trials within a game. The timing of the moves reflected a strong tendency towards higher scoring configurations. For example, in the Same game individuals with zero same neighbors typically moved after less than 2 seconds, while those with 5 or more same neighbors would remain stationary for more than 20 seconds. The motivation to get high scores was also reflected in the students' discussions and exclamations during gameplay. These were primarily about maximizing points, and at no time in any of the trials did any of the students make any wider reference to segregation. Both the actions (in terms of movements made in the game) and verbal expressions were thus consistent with our assumption that the participants saw the game in terms of utility maximization.

Although the participants followed the incentives we assigned them, they produced different outcomes than seen in models. Previous research suggests that this could be due to significantly high levels of behavior noise (8). Our participants inadvertently committed errors but the errors were not the driving mechanism. Instead, it appears that the participants used a strategy that differed from the one implemented in the simulation models. In line with previous work (6-8), our model assumes the best-response strategy, according to which individuals change their position only if it increases their utility. This implies that individuals are able not only to identify better positions but also to recognize when no better positions exist. However, given the fast game dynamics, the participants in our experiment faced cognitive limitations in identifying the optimal locations. In addition, they appeared to be unwilling to "satisfice." The participants moved whenever they did not obtain the perfect score. This is evident from the high mobility in less-than-optimal positions in Figure 4B-D. Moreover, when they moved, the participants did not necessarily move to better positions but rather, appeared to choose their new location randomly (Figure S7). These behavioral rules led to unpredictability and no stable equilibrium arrangement was achieved.

The participants' behavior explains the mismatch with the predictions of the best-response simulation model. In the Same game, the participants obtained more segregated outcomes because they wanted to avoid being on the frontier between the two neighborhoods, as this made them more vulnerable to other participants' moves. In the Same and Diverse game, the participants avoided segregation because they could not be satisfied with being at the periphery of their own-color neighborhood, as these positions entailed lower scores. In the Same or Different game, the participants failed to coordinate on the common-sense solution based on two groupings of yellow and blue because their scores could be easily lowered by one individual of the opposite color infiltrating a mono-color block.

To further test our explanation for the experimental results, we replicated the simulation assuming random relocation and no satisficing. In the new model, agents decide to move whenever their utility is less than the maximum. They then move to a location chosen at random. The predictions from this model match the observed outcomes better, particularly for the Same and Diverse game and the Same or Different game (Figure S8).

**Discussion**

In large cities, people move rental housing and workplaces relatively often, and sometimes without full information about the make up of the new place. Still, residential and work segregation take place on longer time scales, over wider areas, and with much larger costs and benefits than in the experiments we have carried out. We should therefore be careful how we interpret our experimental results in a wider context. Nevertheless, it is important to note that experiments on humans, even on a short time scale, are likely to be more relevant to policy making than simulation results alone. The strong conclusions of (7) and (8), quoted in the introduction, are not supported by our experiment. In the Same and Diverse experiment, segregation is low and those individuals scoring poorly had inadvertently moved in to areas with slightly too high concentrations of dissimilar neighbors. On this basis, these models should not be used as an argument against the need to educate people in the benefits of diversity and we should not conclude that real-world segregation is an unavoidable consequence of weak preferences.

The differences between the experiment and the simulation show how aspects of human psychology that are not captured in simplified simulation models may affect outcomes at the group or society level. The participants in our experiment behaved rationally in the sense that they acted to maximize their utility. Yet, the patterns they created were characteristic to humans rather than simulated agents. Because our experimental games allowed for asynchronous and repeated interactions in a social setting, they were sufficiently flexible to reveal these patterns. The results show that even in a simple setting, humans can act according to the incentives we give them while simultaneously defying our models of what they will do.

**Acknowledgements**

We are grateful for comments from Peter Hedström, Robert D. Mare, Gunn Birkelund, and Michael Mäs. We also thank Michael W. Macy and members of the Social Dynamics Lab at Cornell University for useful feedback.

**Figures and Figure Captions**

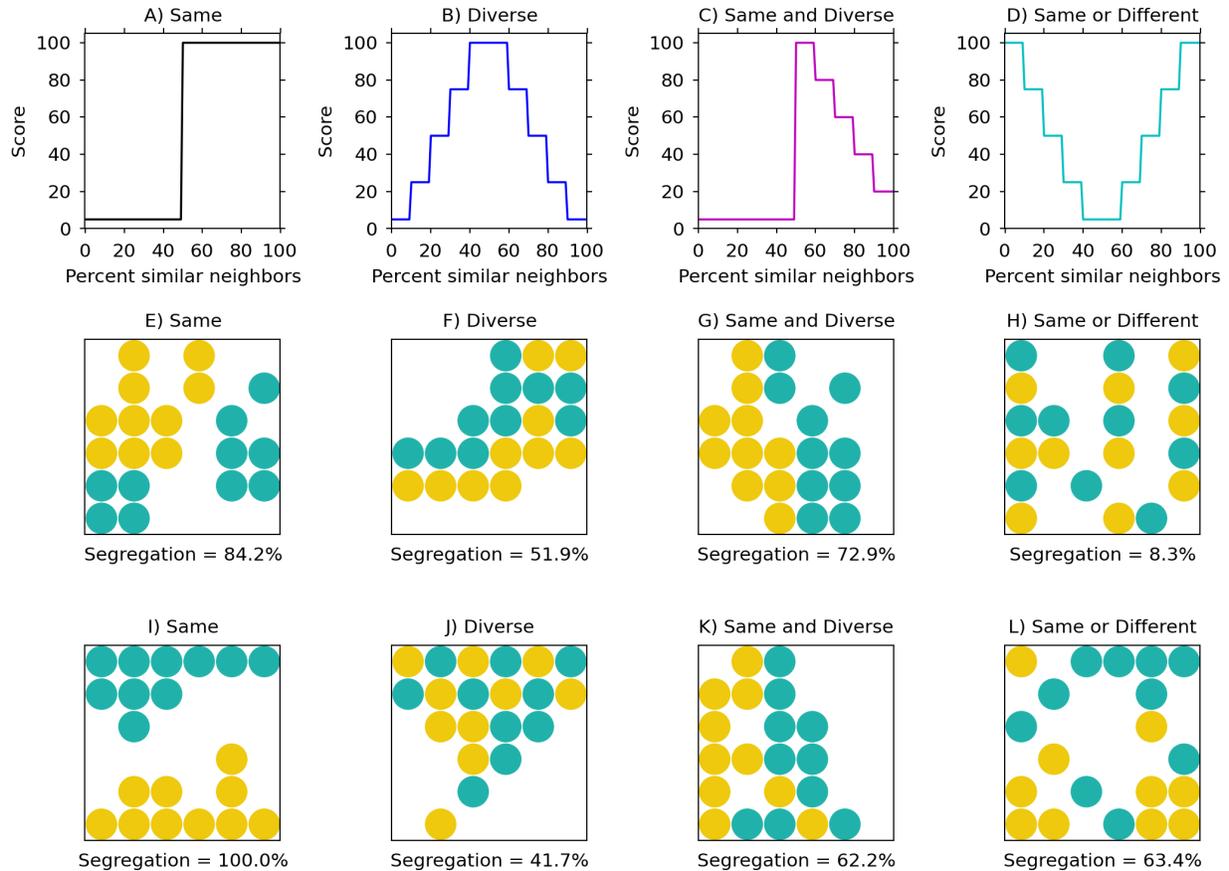

**Figure 1. Utility functions, representative simulation outcomes, and representative experiment outcomes for the four games.** The figure shows the utility functions used in the simulations (and equivalently, the scoring rules in the experiment), a typical outcome in the simulation for a group size of 20, and the outcomes in one of the experimental groups with 20 participants, for the Same game (**A**, **E**, **I**), the Diverse game (**B**, **F**, **J**), the Same and Diverse game (**C**, **G**, **K**), and the Same or Different game (**D**, **H**, **L**).

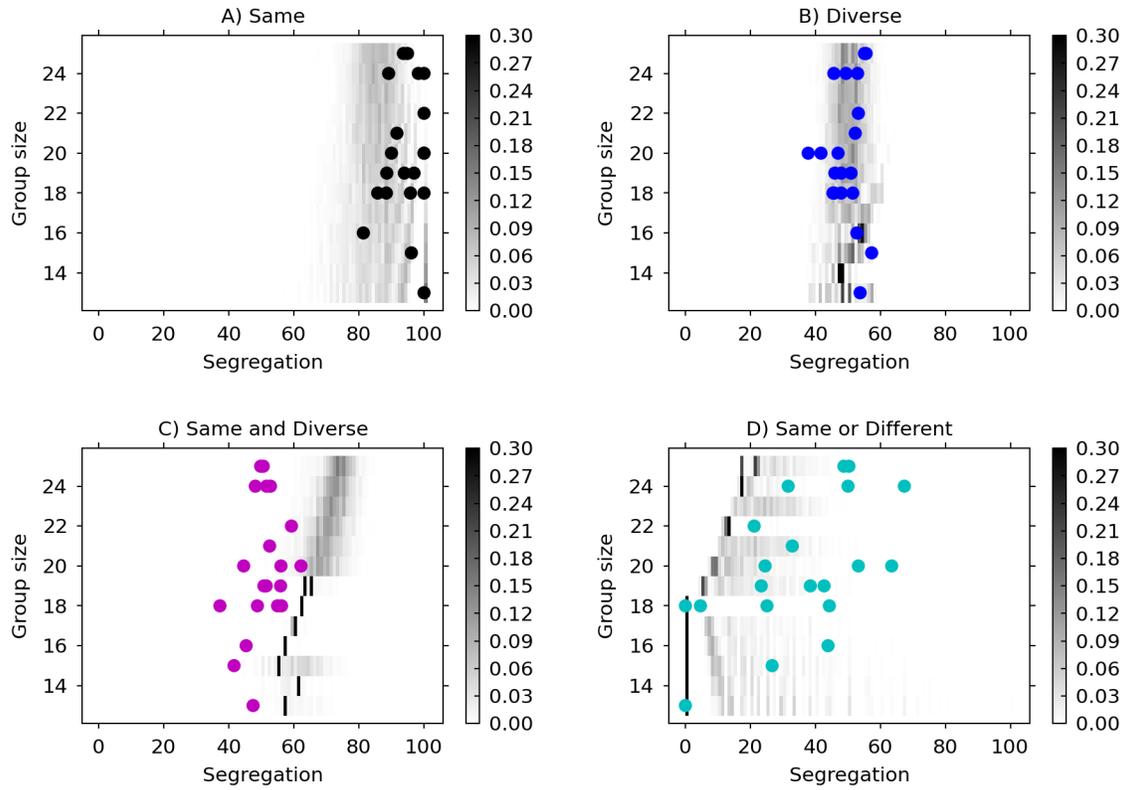

**Figure 2. The segregation reached at the end of each game in the experiment compared to the predictions from the simulations.** Segregation is measured as the average percent similar neighbors. Results are shown for (**A**) the Same game, (**B**) the Diverse game, (**C**) the Same and Diverse game, and (**D**) the Same or Different game.

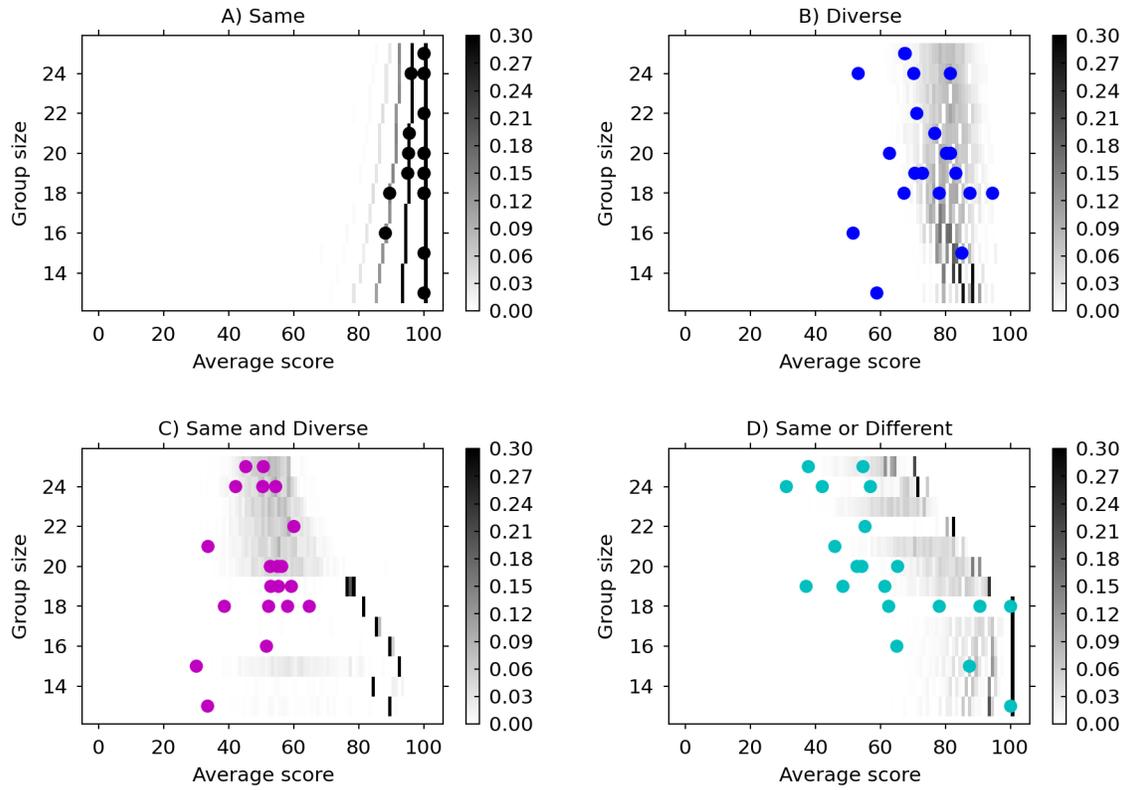

**Figure 3. The average score achieved at the end of each game in the experiment compared to the predictions from the simulations.** Results are shown for (**A**) the Same game, (**B**) the Diverse game, (**C**) the Same and Diverse game, and (**D**) the Same or Different game.

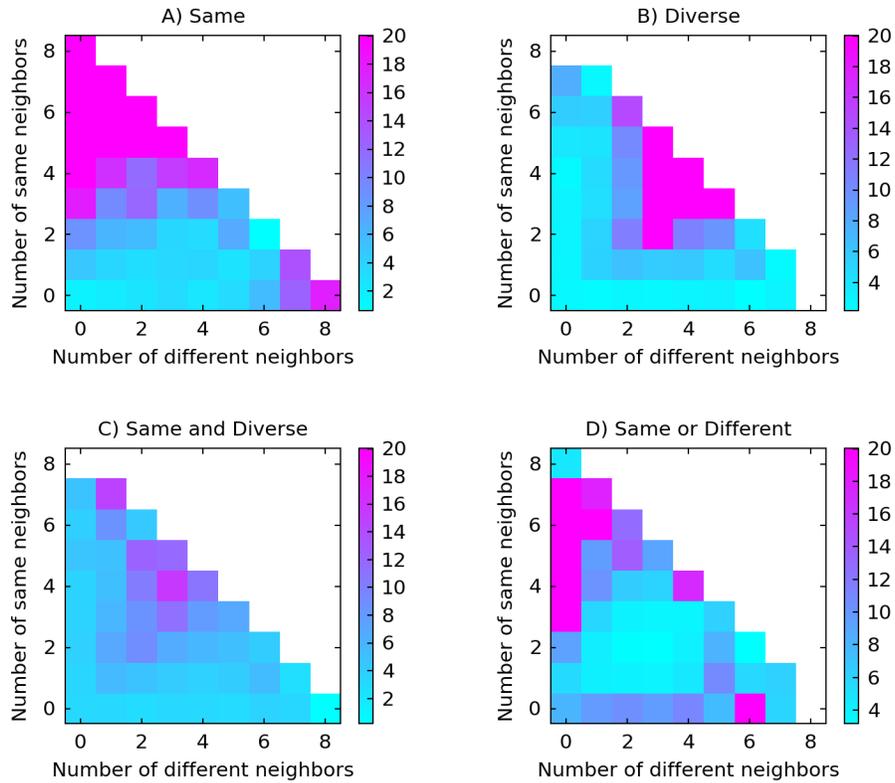

**Figure 4. Average time until next move as a function of the number of different and same neighbors.** Results are shown for (**A**) the Same game, (**B**) the Diverse game, (**C**) the Same and Diverse game, and (**D**) the Same or Different game. Times are estimated by first calculating the total time each neighbor configuration (same, different) is observed over all individuals in all experimental groups. This time is then divided by the number of occurrences of a move away from that configuration.

**An Experimental Study of Segregation Mechanisms**

Milena Tsvetkova[a,1] , Olof Nilsson[b], Camilla Öhman[b], Lovisa Sumpter[c], and David Sumpter[b]

[a] Oxford Internet Institute, University of Oxford, Oxford OX1 3JS, United Kingdom

[b] Department of Mathematics, Uppsala University, 75105 Uppsala, Sweden

[c] School of Education, Health and Social Studies, Dalarna University, 79188 Falun, Sweden

## 1. Simulation

Our simulation to a large extent replicates the model studied by (7). Agents in the model are situated on a two-dimensional 6x6 grid. At every time period, an agent is selected uniformly at random from the population and given the opportunity to relocate if it is unsatisfied. The agents have local information and evaluate the nearest four available locations in the up, down, left and right directions, as well as their current location. If agents identify a single strictly best available location among these, they move to (or stay at) it; if they identify multiple best locations, they randomly choose to relocate to one of them. Thus, as in (7), we assume that agents use a myopic best response, while we also restrict the agents' information and movement to more accurately reflect our experimental design.

To generate predictions for the experiments, we ran 1000 simulations for each utility function in Fig. 1A-D and each population size in the range 13-25. We ran each simulation for 100,000 time periods and recorded the outcome in the last period, regardless of whether a stable equilibrium was reached.

Our results largely replicate the results in (7). We find that segregation obtains even in the presence of a preference for diversity (segregation is similar for the Same and the Same and Diverse utility function) and that integration is possible only when there is no homophily (i.e. for the Diverse utility function; Fig. S1A). In other words, segregation results from the asymmetry of preferences for similar and different neighbors. Segregation obtains for the Same and Diverse utility function because the best-response dynamics eliminate ideal locations quickly and leave only less satisfactory ones. Thus, the dynamics pose a coordination problem – agents need to collectively coordinate on more optimal outcomes, which also happen to be the less segregated equilibria.

Generally, the equilibria for the Same and Diverse utility function are characterized by low average scores (Fig. S1B). Compared to the Same utility function, adding diversity only slightly lowers segregation but at the cost of a large drop in collective happiness. In contrast, preference for diversity without homophily (the Diverse utility function) achieves ideal integration at only slightly lower efficiency than just homophily (the Same utility function).

The Diverse utility function also produces the outcomes with highest clustering (Fig. S1C). While pure homophily (Same) results in a segregated world with clusters with clear boundaries, pure preference for diversity (Diverse) produces tightly knit and well-integrated communities.

The results are largely robust to alternative information and moving rules, as long as the myopic best-response assumption remains. The results hold if we assume global information and global movement, as in (7). They also hold if we assume that agents pick the best location on the whole grid but can only move to the nearest empty spot toward it in one of the four major directions, that is, if we assume global information and local movement. These analyses are not reported here but are available from the authors upon request.

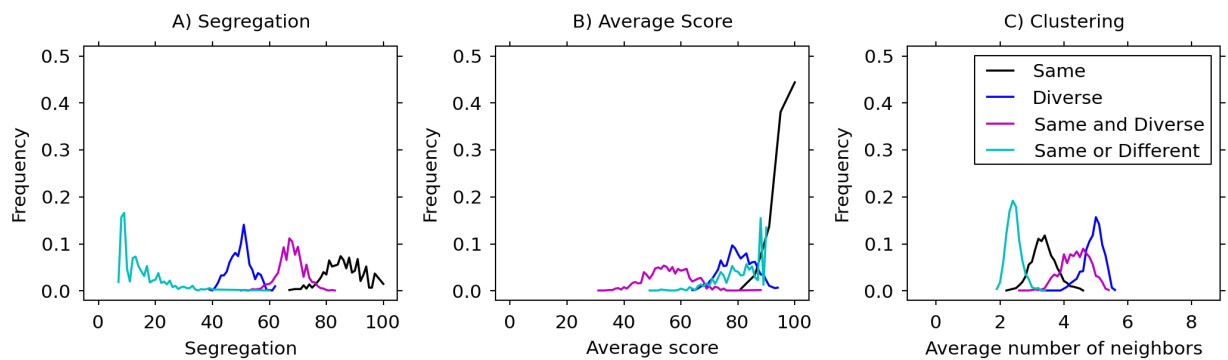

**Fig. S1.** The predictions from the simulations for group size 20. Results are shown for (**A**) average percent of same-color neighbors, (**B**) average score, and (**C**) average number of neighbors.

## 2. Experiment Design

The experiments were conducted as part of an interactive demonstration on mathematical modeling that we presented to high school students in Sweden. The demonstrations aimed to combine disseminating scientific ideas to the general public with collecting data for research. We conducted the sessions in November 2014, December 2014 and March 2015.

Using this population group had a number of advantages and disadvantages. In Sweden, minors in schools do not require parental permission to participate in research conducted during regular class sessions. This age group has almost universal familiarity with computer games and digital devices. This meant that the participants would not only understand the game rules and the game interface more readily but would also be more engaged in playing the game. The latter was particularly important since we were not allowed to use monetary incentives in the school context.

We recruited classes from schools in the Uppland, Väster-götland and Dalarna provinces, which are in the east, the west, and the center of Sweden, respectively. The sample included both urban and rural, and both public and privately managed schools. The schools also largely varied in size. The classes we selected were from non-vocational specializations only, since they tend to have more extensive training in Mathematics. The contacted teachers were instructed not to inform the students what the session would involve, rather to say that researchers from Uppsala and Cornell University would allow them to participate in an experimental game on "modeling social processes."

The games were designed to be as simple and intuitive to play as possible. We restricted movement to a grid, rather than a torus, and to the four main directions, instead of including diagonal movement. We used numbers for the avatars to make them easy to track during the fast game dynamics. We presented the utility function in the form of a score bar, which is a common element in computer games (Fig. S2). Most importantly, we allowed decisions to happen in real time, rather than in periods (Fig. S3). On the one hand, this made the games stimulating and engaging. On the other hand, it helped us avoid forcing the model dynamics on the interactions in the experiment and consequently, prevented us from "cooking in" the theoretical predictions.

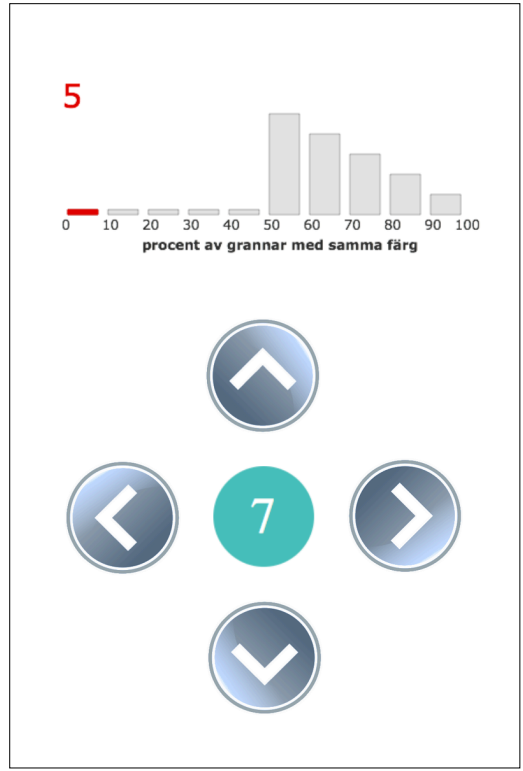

**Fig. S2.** A screenshot of the game control as it was seen on a surf tablet.

We allowed communication during the games as students were in close proximity and full visibility of each other. The play caused a lot of excitement and loud comments, including exclamations of joy for obtaining high score, as well as curses about neighbors who misbehaved. Almost always, there were also students who suggested coordinating strategies to mutually solve the game. For example, in the Same game, someone would yell something similar to "blues to the left, yellows to the right." Surprisingly, this opportunity to coordinate strategies was not always successful. The fact that only three groups managed to coordinate on a checkerboard pattern and zero on a segregated pattern in the Same or Different Game demonstrates this.

At the beginning of the experiment, students' ID numbers were placed face down on their desk or chair. However, due to students being in close physical proximity to each other in the classroom, it was impossible to prevent them from sharing their ID with their neighbors. Further, the ID numbers were in order, as we needed to record the exact seating arrangement both for the demonstrations and for research, and hence students could potentially extrapolate the numbers for everyone else. Nevertheless, there is no evidence that this "weak anonymity" systematically biased participants' behavior (Fig. S4). The games were so fast-paced (3.8 moves per second on average) that it was impossible to keep in mind other players' IDs and colors.

**Fig. S3.** A screenshot of the game, which was projected on a screen at the front of the classroom.

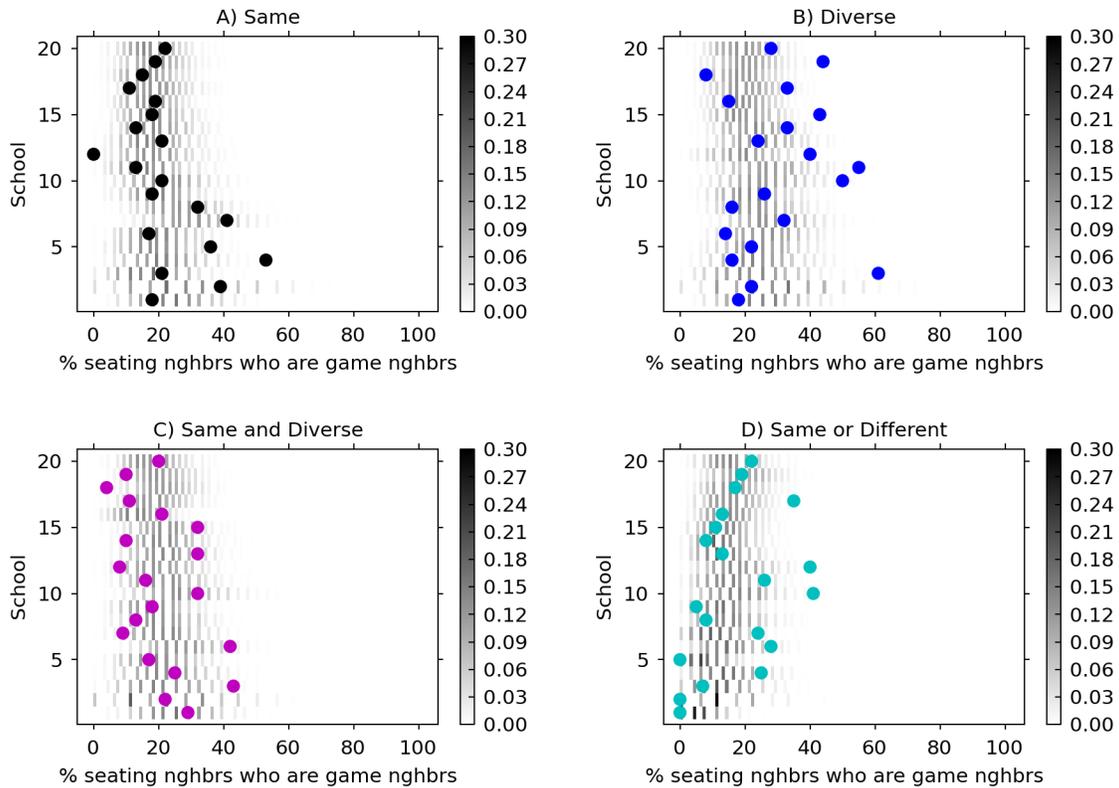

**Fig. S4.** The average percent of participants' two immediate neighbors in the classroom (to the left and right) who were also neighbors at the end of each game in the experiment compared to the expected percent if individuals were randomly assigned to classroom seats. Results are shown for (**A**) the Same game, (**B**) the Diverse game, (**C**) the Same and Diverse game, and (**D**) the Same or Different game.

## 3. Experiment Software

The game platform was developed as a Node.js application, entirely written in JavaScript and HTML 5. The application uses the WebSocket protocol to allow for an ongoing two-way communication between the server and the browser. Data are saved in a MongoDB database, which is a NoSQL database using JSON-like documents with dynamic schemas. All used software is free and open-source and the scripts and communication protocol are compatible with the vast majority of commercial web browsers. The source code for the game platform is available from the authors upon request.

## 4. Experiment Protocol

Students in the classes knew that they would be given an interactive demonstration on modeling a social phenomenon but were not informed what that social phenomenon was. At the beginning of the demonstration, they were informed that their decisions and answers would be recorded for research but kept confidential. The demonstration started with a brief description of the game setup. Then, students logged in the tablets with their assigned ID number, provided their first name, and answered a four-question survey. The survey asked them to indicate their gender (female or male), preferred pet (cat or dog), preferred school subject (Math or Swedish), and preferred leisure activity (computer games or outdoors). This information was used for the demonstration and the data collection process only. Specifically, we deleted the names from the database once we collected all data.

Students then played a trial run and after that, the four games in randomized order. Each time, they were given brief instructions about the score rules in the particular game and then allowed two minutes to play. The score rules were projected on the screen and thus, participants knew that they were all playing the same game. After the games, students were asked to complete a short survey on their national origin, academic aspirations, family socioeconomic status, and social network. Then, their final scores were revealed in a list on the screen.

At that moment, we revealed that the games were based on a mathematical model of segregation. We then gave a lecture on the problem of segregation and the Schelling model. In the lecture, we used the students' answers from the first four-question survey and the final configurations in the games to demonstrate the problem of segregation and the model. Additionally, the demonstration included a ten-minute segment in which students were asked to run simulations of the model (using a web application we had developed) in order to answer a few questions about the model behavior. The introduction and the games were lead in Swedish, while the lecture was given in English.

The following is the original English version of the demonstration protocol. This version was translated into Swedish for the experiments:

## Segregation in Minutes

"Segregation in Minutes" is an educational demonstration and behavioral study that uses a web-based platform and an online real-time multi-player game. The session is introduced to students as an interactive lesson on mathematical modelling. To prevent behavioral and response bias, it is important not to reveal the topic of segregation to students before all data have been collected (see below).

The goal of the session is twofold:

1) to demonstrate basic concepts of segregation to students according to the Schelling model;
2) to collect data on individual and collective behavior for research.

### Required Equipment

- 25 surf tablets with a browser with capability for web sockets to serve as game "consoles" for each student
- Instructor's laptop
- 36 ID cards, numbered from 1 to 36 (the login ID)
- Overhead projector and screen

### Method

1) Setup

✎ Arrange the seats in the classroom on a 6x6 square grid. On each seat, place one of the ID cards <u>facing down</u>. Arrange the cards so that the numbering increases consecutively from left to right and then from top to bottom:

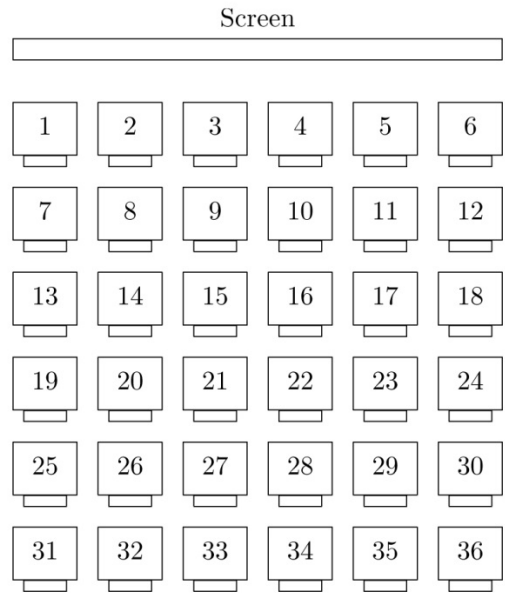

✎ Choose (uniquely) one of the orders in which Game 1, Game 2, Game 3, and Game 4 will be played in the session.

| | | | |
|---|---|---|---|
| 1, 2, 3, 4 | | 3, 1, 2, 4 | |
| 1, 2, 4, 3 | | 3, 1, 4, 2 | |
| 1, 3, 2, 4 | | 3, 2, 1, 4 | |
| 1, 3, 4, 2 | | 3, 2, 4, 1 | |
| 1, 4, 2, 3 | | 3, 4, 1, 2 | |
| 1, 4, 3, 2 | | 3, 4, 2, 1 | |
| 2, 1, 3, 4 | | 4, 1, 2, 3 | |
| 2, 1, 4, 3 | | 4, 1, 3, 2 | |
| 2, 3, 1, 4 | | 4, 2, 1, 3 | |
| 2, 3, 4, 1 | | 4, 2, 3, 1 | |
| 2, 4, 1, 3 | | 4, 3, 1, 2 | |
| 2, 4, 3, 1 | | 4, 3, 2, 1 | |

☚ Log in to the site as admin and set up the session.

2) Students enter

💬 *Welcome. Please take a seat. You can sit wherever you would like.*

3) (3 min) Introduction

💬 *I will start by reading you the instructions for this session.*
💬 *You are about to participate in an interactive demonstration on mathematical modeling. (Mathematical modeling is about describing events with the help of mathematics.) First, you will play 4 different games. The tablets will serve as game controls; the game itself will be projected on the screen. Then, you will be asked to fill out a short survey. Finally, you will explore an interactive simulation model on which the games are based.*
💬 *At the beginning, we will ask you for your name but we will need it only for the demonstration. Eventually, we will anonymize all information and your name will be deleted. Your answers and decisions will be saved for research on mathematical modeling.*

4) (7 min) General game instructions

💬 *As I mentioned, we will start by playing 4 games. The games have different scoring rules but in each game, your goal is to score the highest number of points at the end of the game.*
💬 *In each game, your avatar is shown by a colored circle with a number. Your avatar will be one of two colors: either yellow or blue. All avatars are situated on a grid.*
💬 *On the grid, you can move your avatar left, right, up, or down to the next empty spot in the specified direction, if such a spot exists. If the spot immediately next to you is occupied by another avatar, you will jump to the next empty spot. If you are at the edge of the grid and there are no more spots in the direction you want to move, you will not move.*
💬 *In each game, your score depends on the color of your neighbors. A neighbor is anyone who occupies one of the 8 cells surrounding you. Thus, you can have at most 8 neighbors. However, if there are empty spots around you, you will have fewer than 8 neighbors.*
💬 *You can use your game control to move your avatar and track your score. The heights of the score bars show how many points you get for different percent of same-color neighbors. The red bar and the red text show your current score.*
💬 *The games have different scoring rules but in each game, your goal is to score the highest number of points at the end of the game. Each game will end once everyone is satisfied with their location and all movement has stopped, or after 2 minutes of play at most.*
💬 *After we are done with all 4 games, we will display each person's total score on the screen.*

💬 *Are there any questions?*

☙ Distribute tablets to all students. Describe how to handle the device.

💬 *To turn the tablet on, press the button on the corner of one of the long sides. To go to the desktop, press the arrow in the center at the very bottom of the screen. Please try not to press any other buttons or icons because that will interrupt your game. Feel free to rotate the tablet if you cannot view all of the information on the screen.*

💬 *To log in, please click on the Play icon on the tablet screen. Please log in to the site with the ID written on the card on your desk. Then, please submit the additional login information. You will then reach the game control. However, we need to wait for everyone to be done before you can use the game control.*

5) (4 min) Trial Run

☙ Wait for all students to complete the survey.

💬 *Before you play the games, we will do a trial run so that you can practice how to use the game control and how to follow your avatar on the screen. Your score from the trial run will not be used for calculating your final score. The trial run is intended for practice only.*

💬 *Your score will be between 5 and 100 depending on the percent of same-color neighbors you have and the particular game you are playing. Each score bar corresponds to different percent of same-color neighbors: between 0 and 9 percent, between 10 and 19 percent, and so on up to 100 percent. The height of each score bar shows how many points you get if you have that percent of same-color neighbors. The red bar corresponds to your current neighborhood and the red text shows your current score.*

☙ Select [ Trial Run ▾ ] from the drop-down menu and [ Initialize ] . Read the Trial Run instructions. [ Start game ] .

6) (4 x 3 min) Games

💬 *Now, we will play the 4 games. We will go over the scoring rules before each game. Remember, the score you obtain at the end of each game will be used to calculate your total score. At the end, we will display each person's total score on the screen.*

☙ For each of the four games: Select the game ( [ Game 1 ▾ ] , [ Game 2 ▾ ] , etc.) according to the predetermined random order. [ Initialize ] . Read the game instructions. [ Start game ] .

7) (5 min) Survey

💬 *Before we show you your final score, we would like to ask you to answer a short survey that we need for research. Your answers will be kept confidential and not shared with anyone who is not a researcher in our study. When you have completed the survey, please leave the tablet in front of you on your desk.*

8) (1 min) Final Scores

Click on  Final scores  to show student scores.

9)  (10 min) Lecture: Modeling Segregation

*These games were designed to demonstrate to you the social phenomenon of* **segregation.**
   ➢ Show lecture slides 1–19.
Use simulation model to demonstrate segregation outcome in mathematical model.

   Display  results for  Game 1 Start ▾  compared to  Game 1 End ▾ .

   ➢ Show lecture slides 20–21.
Display  results for  Game 1 End ▾  compared to segregation in current seating
   arrangement as per the  Survey ▾ . Use  Previous  and  Next  to cycle
   through the different survey answers.
   ➢ Show lecture slide 22.

10) (12 min) Simulation model

Distribute task sheets to students.
   ➢ Show lecture slide 23.
*You now have 10 min to use the simulation model and study what happens if people prefer both similarity and diversity. You can work in pairs or in groups of 3.*
*To go to the simulation, please click on the Simulation icon on the tablet screen.*

11) (5 min) Lecture: Segregation and Integration

*What did you find out?*
Display  results for  Game 1 End ▾  compared to  Game 2 End ▾ .
Display  results for  Game 1 End ▾  compared to  Game 3 End ▾ .
   ➢ Show lecture slides 24–26.

**Legend**
To do
To say
   ➢ To show from lecture slides

## 5. Additional Analyses

### 5.1. Time Series

Our analysis of the game dynamics suggests that most of the experimental groups converged to a dynamic equilibrium within the allowed two minutes of play. The convergence was remarkably fast in the Same game (Fig. S5A and S5E) and relatively gradual in the Diverse game (Fig. S5F). There was very little change in the levels of segregation and the average scores in the Same and Diverse game (Fig. S5C and S5G), while the change in the levels of segregation in the Same or Different game shows a shift in strategies: an initial attempt at the "Same" strategy followed by the "Different" strategy. The latter game also exhibits the largest variation in dynamics and outcomes, similarly to the simulation predictions.

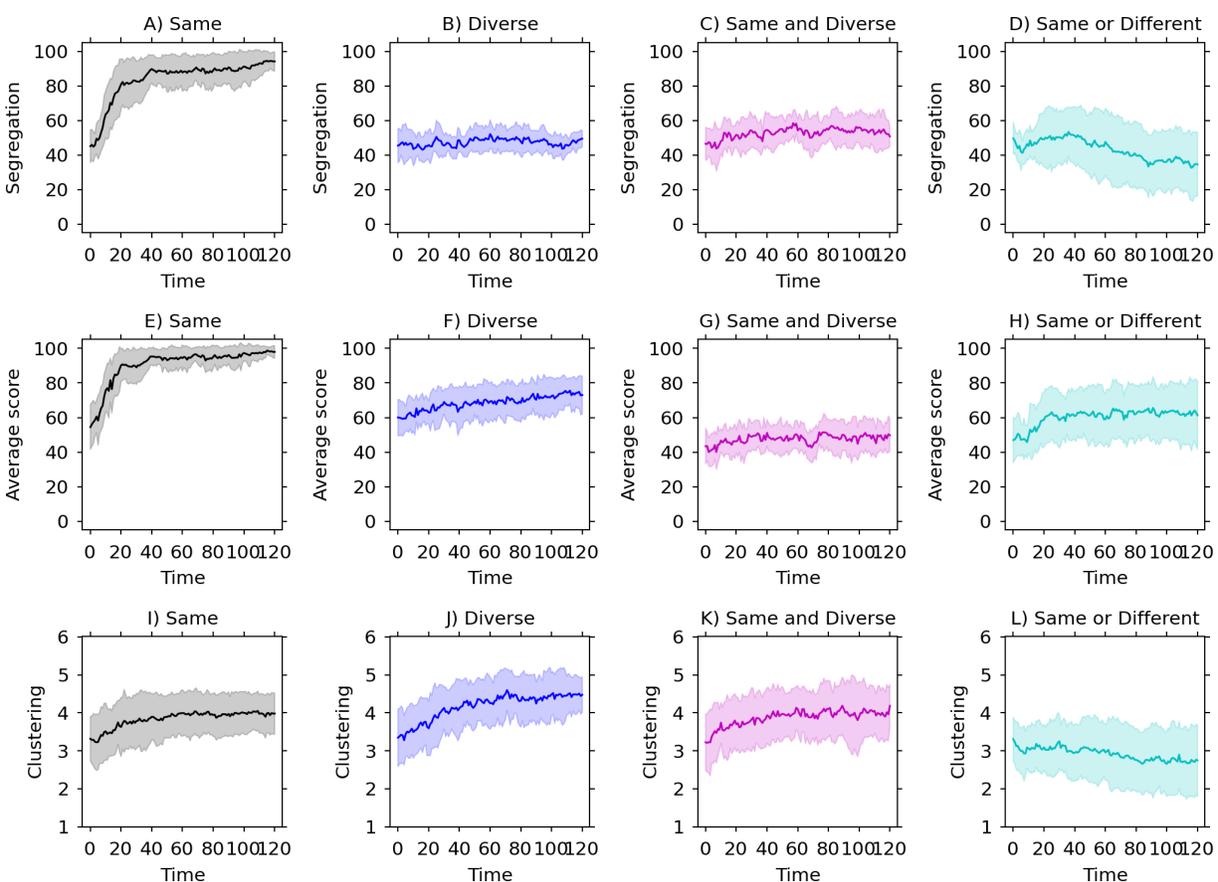

**Fig. S5.** Time series of the level of segregation, the average score, and the average number of neighbors in the experiment. The figures show the mean and one standard deviation from the mean for the Same game (**A**, **E**, **I**), the Diverse game (**B**, **F**, **J**), the Same and Diverse game (**C**, **G**, **K**), and the Same or Different game (**D**, **H**, **L**).

### 5.2. Clustering

Overall, the experimental groups achieved clustering that is similar to the clustering predicted by the simulations. There was only a slight tendency for the Same game to end more clustered than predicted and the Diverse game to be less clustered than predicted (Fig. S6).

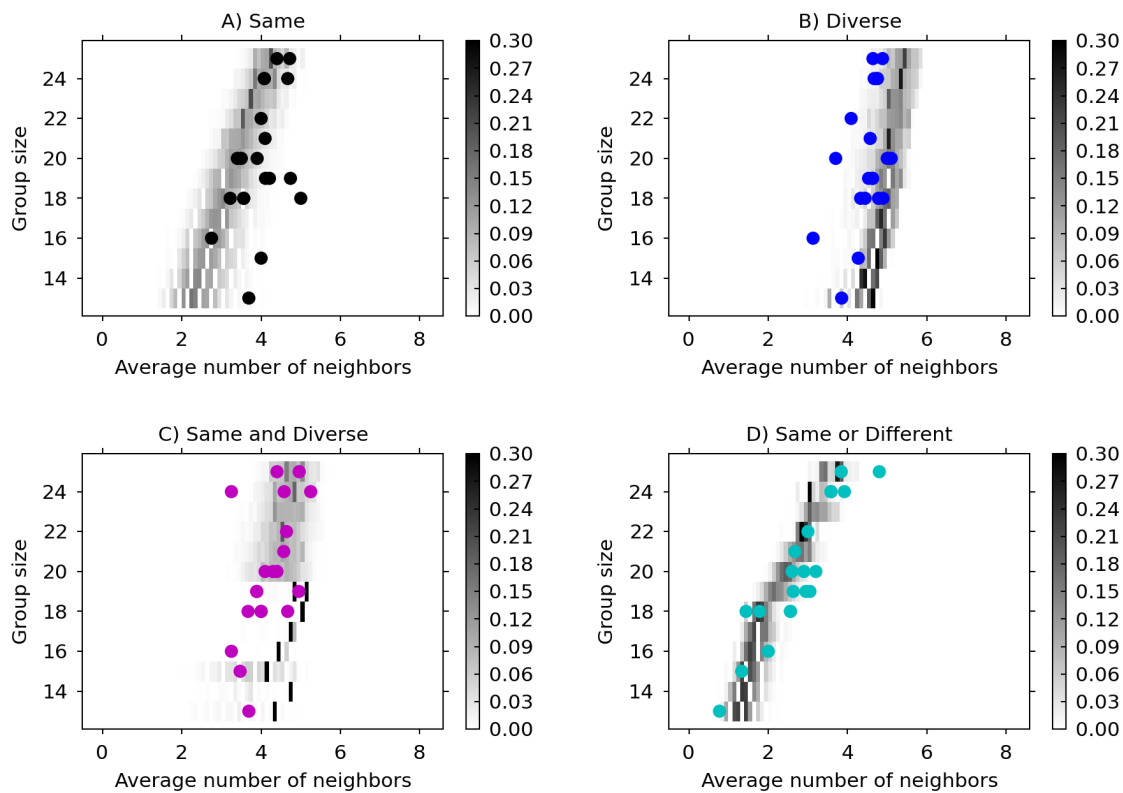

**Fig. S6.** The average number of neighbors at the end of each game in the experiment compared to the predictions from the simulations. Results are shown for (**A**) the Same game, (**B**) the Diverse game, (**C**) the Same and Diverse game, and (**D**) the Same or Different game.

### 5.3 Movement Strategies

We investigated whether the moves were score-maximizing by plotting the percent of moves that originated in a certain score that resulted in another score. If participants moved mainly to locations that increased their scores, we would expect a higher number of moves in the right half of the plots on Fig. S7. Except for the Same game, this is clearly not the case. It appears that participants chose a random direction when they decided to move. There is no evidence that they preferred the nearest empty spot that increased their scores.

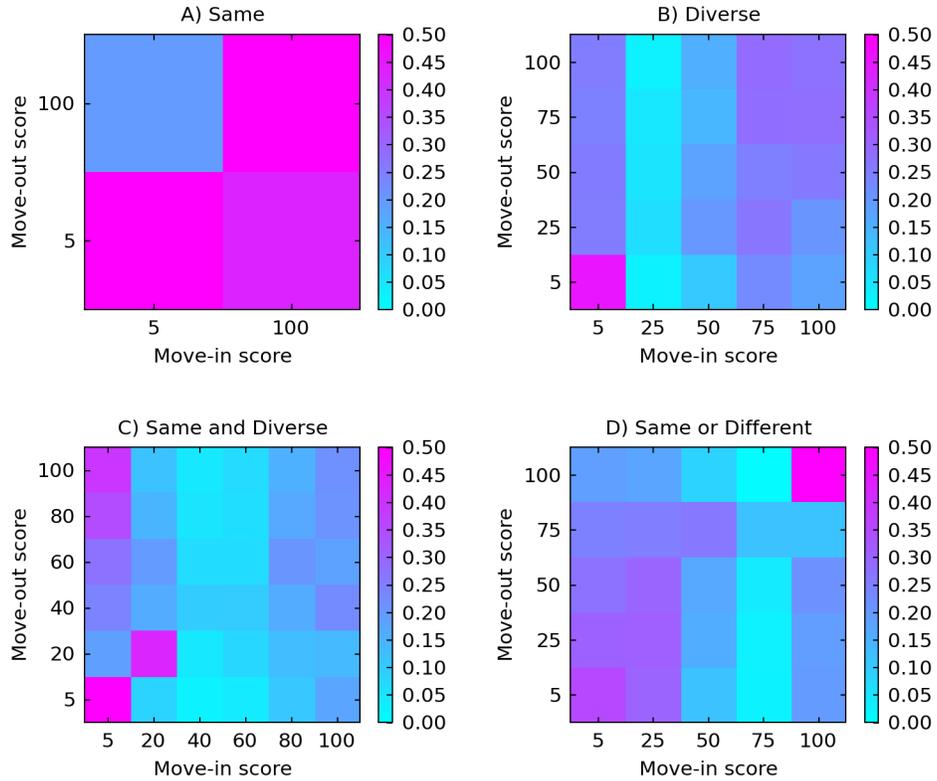

**Fig. S7.** The frequency of score changes as a result of a move. Results are shown for (**A**) the Same game, (**B**) the Diverse game, (**C**) the Same and Diverse game, and (**D**) the Same or Different game. The frequencies are normalized along each row. Each cell shows the proportion of moves away from a particular score (move-out score) that result in the move-in score.

### 5.4. Simulation with Random Relocation

To provide a better explanation for the experimental results, we replicated the simulation without the best-response assumption. In the new version, agents decide to relocate when they do not obtain the maximum utility at their current location and move to one of the nearest four available locations in the up, down, left and right directions chosen at random. The predictions from this model match the observed outcomes significantly better (Fig. S8).

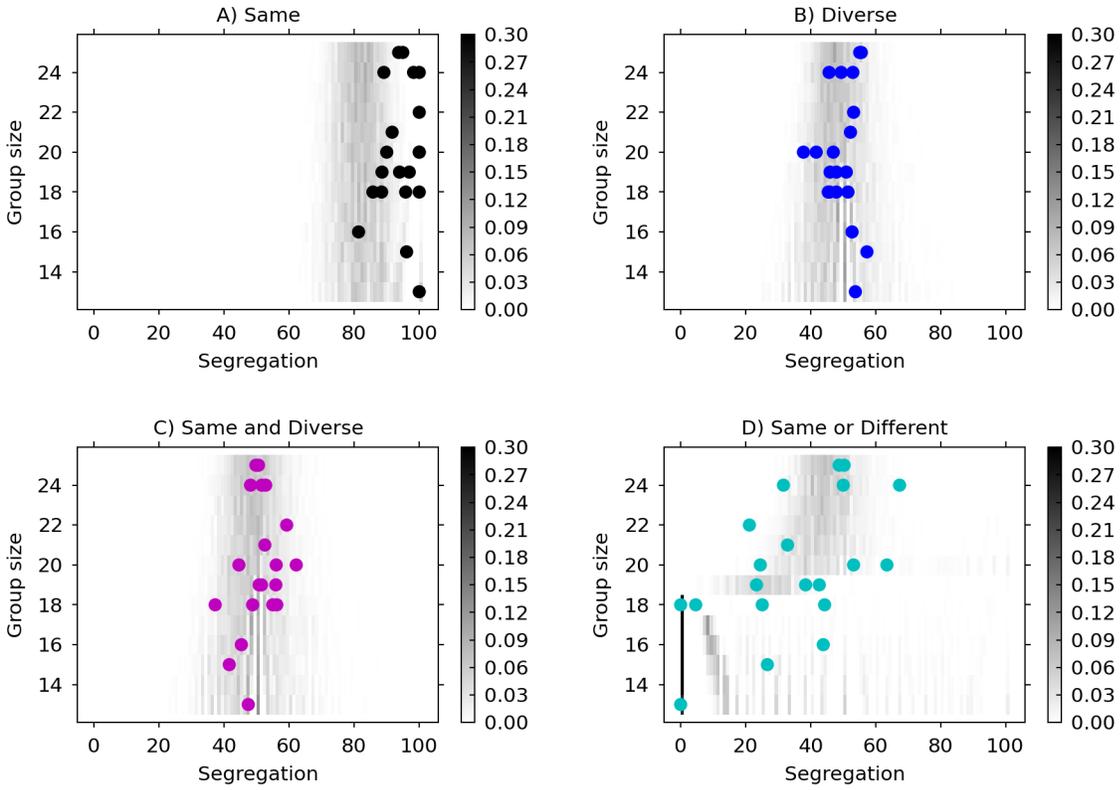

**Fig. S8.** The segregation reached at the end of each game in the experiment compared to the predictions from a simulation model with random relocation. Segregation is measured as the average percent similar neighbors. Results are shown for (**A**) the Same game, (**B**) the Diverse game, (**C**) the Same and Diverse game, and (**D**) the Same or Different game.